\documentstyle[preprint,tighten,aps]{revtex}
\begin{document} 
\draft
\preprint{IASSNS-HEP-96/43,UPR-700-T,PUPT-1623}
\date{May 1996}
\title{Near-BPS-Saturated Rotating Electrically Charged Black Holes 
as String States}
\author{Mirjam Cveti\v c$^1$
\thanks{On sabbatic leave from the University of Pennsylvania.
E-mail address: cvetic@sns.ias.edu}
and Donam Youm$^2$
\thanks{On leave from the University of Pennsylvania. 
E-mail addresses: youm@pupgg.princeton.edu; youm@sns.ias.edu}}
\address{$^1$ School of Natural Science, Institute for Advanced
Study\\ Olden Lane, Princeton, NJ 08540 \\ and \\
$^2$ Physics Department, Joseph Henry Laboratories\\ 
Princeton University, Princeton, NJ 08544}
\maketitle
\begin{abstract}
{We construct generating solutions for general $D$-dimensional 
($4\le D\le 9$) rotating, electrically charged, black holes in the 
effective action of toroidally compactified heterotic (or Type IIA) 
string.  The generating solution is parameterized by the ADM mass, 
two electric charges and $\left[{D-1}\over 2\right]$ angular momenta 
(as well as the asymptotic values of one toroidal modulus and the 
dilaton field).  For $D\ge 6$, those are generating solutions 
for {\it general} black holes in toroidally compactified  heterotic  
(or type IIA) string. Since in the BPS-limit (extreme limit) these 
solutions have singular horizons or naked singularities, we 
address the near extreme solutions with all the angular momenta 
small enough.  In this limit, the thermodynamic entropy can be 
cast in a suggestive form, which has a qualitative interpretation 
as microscopic entropy of (near)-BPS-saturated charged string 
states of toroidally compactified heterotic string, whose 
target-space angular momenta are identified as $[{{D-1}\over 2}]$ 
$U(1)$ left-moving world-sheet currents.}
\end{abstract}

\section{Introduction}

Recently, black holes in string theory have become a  subject of  
intense research since it has now become possible to address their 
quantum properties, and in particular the microscopic origin of the 
entropy for certain classes of black holes.  In particular, for 
certain five-dimensional [four-dimensional] BPS-saturated static 
\cite{SV,CM,DVV} [\cite{MSTR,JKM}] and rotating \cite{BMPV} 
[\cite{HLM}], and near-extreme static \cite{DAS,CM,HS} and rotating 
\cite{BLMPSV,Maldacena} [\cite{HLM}] black holes, which can be 
identified with particular $D$-brane configurations, their 
microscopic entropy can be calculated by applying the 
``$D$-brane technology'' \cite{CJP}.
For a recent review of recent  developments in the black hole 
physics in string theory see Ref.\cite{Horowitz}.  For some 
related approaches either from the point of view of M-theory and  
other pure $D$-brane configurations see \cite{Tseytlin,CS,KT,BL}.

An earlier complementary approach to calculate the 
microscopic entropy of four-dimensional BPS-saturated black holes  
was initiated in Ref. \cite{LW} and was further elaborated on 
in Refs. \cite{CT,T} and \cite{LWI}.  These approaches identify 
the microscopic black hole degrees of freedom as quantum hair 
\cite{LW} associated with particular small scale (marginal) 
perturbations \cite{CT,LWI} of string theory, which does not 
change the large scale properties of the black hole solutions.  
On the one hand, the magnetic charges of such dyonic black 
hole solutions ensure that the classical solutions have regular 
horizons \cite{KLOPV,CY} (and thus $\alpha^{\prime}$ corrections 
are under control), while on the other hand they effectively 
renormalize \cite{LW,CT,T,LWI} the string tension of the 
underlying string theory. 
A large part of these ideas was build on an earlier proposal of Sen 
\cite{Sen} (and further explored in Refs. \cite{Peet,CMP,DGHW}) to 
identify the area of BPS-saturated electrically charged spherically 
symmetric solutions, evaluated at the {\it stretched} horizon 
\cite{Susskind}, as the microscopic entropy associated with the 
elementary BPS-saturated states of string  theory.  Since the 
stretched horizon is determined up to ${\cal O}
(\sqrt{\alpha^{\prime}})$, the identification of the two 
quantities agrees up to ${\cal O}(1)$, only.

In this paper we would like to amplify the idea \cite{Sen} to 
identify the  near-BPS-saturated electrically charged black 
holes in string theory with elementary string excitations, 
by addressing such near-BPS-saturated {\it rotating} black holes.  
In particular, we would like to propose to trade the evaluation of 
the area of the BPS-saturated rotating solution at the 
stretched horizon for  the evaluation of {\it the area of the 
horizon of the near-BPS-saturated rotating black hole} and 
equate the latter quantity with the microscopic entropy of 
the (near)-BPS-saturated charged string states with non-zero 
angular momenta.  In this proposal, the role of the distance  
between the stretched and the (singular) horizon of the 
BPS-saturated states is traded for the non-extremality 
parameter, which parameterizes a deviation of the 
near-BPS-saturated solution from the BPS-saturated one.

With that in mind we construct the generating solution for the 
general rotating  electrically charged solution of toroidally 
compactified heterotic (or Type IIA) string in $D$-dimensions 
($4\le D\le 9$).  The generating solution is parameterized by the 
ADM mass $M_{BH}$ (or equivalently, by a non-extremality 
parameter $m$, the mass of the corresponding Kerr solution), 
two electric charges $Q^{(1),(2)}_1$ (associated with the 
Kaluza-Klein and the two-form $U(1)$ gauge fields of one 
compactified direction) and $\left[{D-1}\over 2\right]$ 
angular momenta $J_i$, $i=1,...,\left[{{D-1}\over 2}\right]$, 
(as well as the asymptotic values of one toroidal modulus 
$G_{11\,\infty}$ and the dilaton field $\varphi_\infty$).  
For $D\ge 6$, the general rotating black hole solutions can 
have only electric charges, and thus for $D\ge 6$ these 
solutions are the generating solutions for {\it general} black 
holes of toroidally compactified heterotic (or type IIA) string.

This construction thus adds to a completion of a program 
to obtain the general {\it rotating} black hole solutions of 
toroidally compactified heterotic (or Type IIA) string in 
$4\le D\le 9$.  
The generating solution for the general rotating five-dimensional 
solutions (parameterized by the ADM mass, two angular momenta and 
{\it three} charges) were given in \cite{CYII,CYIII}, while the 
four-dimensional rotating generating solution is missing one more 
charge parameter, {\it  i.e.}, the generating rotating solution  
obtained up to now is in terms the ADM mass, the angular momentum 
and four (two electric and two magnetic) charge parameters 
\cite{CYIII}, only. 
In the BPS-limit (extreme limit), these solutions have at most 
singular horizons, and thus, they are not black holes in the 
conventional sense.   Consequently, for $D\ge 6$ there are no 
BPS-saturated solutions with regular horizons.
Similar conclusions for spherically symmetric, static, 
BPS-saturated solutions were obtained in \cite{KT,pc} by considering 
a general sigma model with $O(D-1)$ spherical symmetry.

In the analysis of the classical solutions, we shall concentrate on 
near-BPS-saturated rotating solutions with the {\it macroscopic} 
values of the charges and angular momenta.  For that purpose,  
$Q_1^{(1),(2)}\gg  m={\cal O}(1)$ (measured in units of 
$\alpha^{\prime}$).  In addition, the inequalities 
$Q_1^{(1)}Q_1^{(2)}\gg J^2_{1,\cdots,[{{D-1}\over 2}]}\gg 
{\sqrt{ Q_1^{(1)}Q_1^{(2)}}}$ ensure that on the one hand  
the solution has regular horizons, and on the other hand that 
the contribution of the angular momenta to the entropy is still  
non-negligible macroscopically. 

We then calculate the degeneracy of states in the 
(near)-BPS-saturated string states with the same target 
space values of the physical parameters.   For that purpose  
the target-space angular momenta of the string states  
are identified with  $[{{D-1}\over 2}]$ $U(1)$ left-moving 
world-sheet currents.  The logarithm of the degeneracy of 
such states is in qualitative agreement with the thermodynamic 
entropy of the near-BPS-saturated black holes, thus lending a 
support to the identification of the microscopic degrees 
of freedom of such black holes with the elementary string 
excitations.

The paper is organized in the following way.
In Section II, the explicit form of the generating solution for the  
general $D$-dimensional rotating electrically charged black hole 
solution of toroidally compactified string is given.  We address the 
singularity structure and write down the thermodynamic entropy for
such near-BPS-saturated black holes.  In Section III, we calculate 
the degeneracy of states of (near)-BPS-saturated string states with 
the same quantum numbers and obtain a  (qualitative) agreement 
between the microscopic (statistical) and  macroscopic (thermodynamic) 
entropies of these black holes.

\section{$D$-Dimensional General Rotating Charged Solution}

We shall present an explicit form of the (generating) solution for 
the general rotating electrically charged black hole of 
$D$-dimensional ($4\le D\le 9$) toroidally compactified heterotic  
(or type IIA) string.  We choose to parameterize the generating 
solution in terms of massless fields of the heterotic string 
compactified on a $(10-D)$-torus (or Neveu-Schwarz-Neveu-Schwarz 
(NS-NS) sector of the Type IIA string compactified on a 
$(10-D)$-torus).  
A subset of $T$-duality symmetry transformations (which do not 
affect the $D$-dimensional space-time) allows one to obtain the most 
general rotating electrically charged solutions in this class.  

Note also that the string-string duality relates these solutions to  
solutions of other string vacua.

\subsection{Effective Action of Heterotic String on Tori}

We briefly summarize the results of the effective action of 
toroidally compactified heterotic string in $D$-dimensions, 
following  Refs. \cite{MSCH,SENFOUR}.  This subsection is  
essentially the same as the one of \cite{CYIII}, however, we 
include it in this paper for the sake of completeness. 

The compactification of the extra $(10-D)$ spatial coordinates 
on a $(10-D)$-torus can be achieved by choosing the following 
Abelian Kaluza-Klein Ansatz for the ten-dimensional metric
\begin{equation}
\hat{G}_{MN}=\left(\matrix{e^{a\varphi}g_{{\mu}{\nu}}+
G_{{m}{n}}A^{(1)\,m}_{{\mu}}A^{(1)\,n}_{{\nu}} & A^{(1)\,m}_{{\mu}}
G_{{m}{n}}  \cr  A^{(1)\,n}_{{\nu}}G_{{m}{n}} & G_{{m}{n}}}\right),
\label{4dkk}
\end{equation}
where $A^{(1)\,m}_{\mu}$ ($\mu = 0,1,...,D-1$; 
$m=1,...,10-D$) are $D$-dimensional Kaluza-Klein $U(1)$ gauge fields,  
$\varphi \equiv \hat{\Phi} - {1\over 2}{\rm ln}\,{\rm det}\, G_{mn}$ 
is the $D$-dimensional dilaton field, and $a\equiv {2\over{D-2}}$.   
Then, the affective action is specified by the following massless 
bosonic fields: the (Einstein-frame) graviton $g_{\mu\nu}$, the 
dilaton $e^{\varphi}$, $(36-2D)$ $U(1)$ gauge fields ${\cal A}^i_{\mu}
\equiv (A^{(1)\,m}_{\mu},A^{(2)}_{\mu\,m},A^{(3)\,I}_{\mu})$ defined 
as $A^{(2)}_{\mu\,m} \equiv \hat{B}_{\mu m}+\hat{B}_{mn}
A^{(1)\,n}_{\mu}+{1\over 2}\hat{A}^I_mA^{(3)\,I}_{\mu}$, 
$A^{(3)\,I}_{\mu} \equiv \hat{A}^I_{\mu} - \hat{A}^I_m 
A^{(1)\,m}_{\mu}$, and the following symmetric $O(10-D,26-D)$ 
matrix of the scalar fields (moduli):
\begin{equation}
M=\left ( \matrix{G^{-1} & -G^{-1}C & -G^{-1}a^T \cr 
-C^T G^{-1} & G + C^T G^{-1}C +a^T a & C^T G^{-1} a^T 
+ a^T \cr -aG^{-1} & aG^{-1}C + a & I + aG^{-1}a^T} 
\right ), 
\label{modulthree}
\end{equation} 
where $G \equiv [\hat{G}_{mn}]$, $C \equiv [{1\over 2}
\hat{A}^{(I)}_{{m}}\hat{A}^{(I)}_{n}+\hat{B}_{mn}]$ and 
$a \equiv [\hat{A}^I_{{m}}]$ are defined in terms of the 
internal parts of ten-dimensional fields.  Then the 
$D$-dimensional effective action takes the form:
\begin{eqnarray}
{\cal L}&=&{1\over {16\pi G_D}}\sqrt{-g}[{\cal R}_g
-{1\over {D-2}}g^{\mu\nu}\partial_{\mu}\varphi
\partial_{\nu}\varphi+{1\over 8}g^{\mu\nu}
{\rm Tr}(\partial_{\mu}ML\partial_{\nu}ML)-{1\over{12}}
e^{-2a\varphi}g^{\mu\mu^{\prime}}g^{\nu\nu^{\prime}}
g^{\rho\rho^{\prime}}H_{\mu\nu\rho}H_{\mu^{\prime}\nu^{\prime}
\rho^{\prime}} \cr
&-&{1\over 4}e^{-a\varphi}g^{\mu\mu^{\prime}}g^{\nu\nu^{\prime}}
{\cal F}^{i}_{\mu\nu}(LML)_{ij}
{\cal F}^{j}_{\mu^{\prime}\nu^{\prime}}],
\label{effaction}
\end{eqnarray}
where $g\equiv {\rm det}\,g_{\mu\nu}$, ${\cal R}_g$ is the Ricci 
scalar of $g_{\mu\nu}$, and ${\cal F}^i_{\mu\nu} = \partial_{\mu} 
{\cal A}^i_{\nu}-\partial_{\nu} {\cal A}^i_{\mu}$ are the 
$U(1)^{36-2D}$ gauge field strengths.  

We take the units such that $c={h\over {2\pi}}=1$.  
The $D$-dimensional 
gravitational constant $G_D$ is related to the 10-dimensional 
gravitational constant $G_{10}$ as:
$G_{10} = ({\rm volume\ of\ T^{10-D}})\times G_D$. 
The canonical choice of the asymptotic values of scalar 
fields, {\it i.e.},  $G_{ij\infty}=\delta_{ij}$ (and 
$\varphi_\infty=0$), corresponds to the compactification on  
$(10-D)$-self-dual circles with radius $R=\sqrt{\alpha^{\prime}}$.   
In this case, $G_D =G_{10}/(2\pi R)^{10-D}= G_{10}/(2\pi\sqrt
{\alpha^{\prime}})^{10-D}$. 

The $D$-dimensional effective action (\ref{effaction}) is 
invariant under the $O(10-D,26-D)$ transformations ($T$-duality) 
\cite{MSCH,SENFOUR}:
\begin{equation}
M \to \Omega M \Omega^T ,\ \ \ {\cal A}^i_{\mu} \to \Omega_{ij}
{\cal A}^j_{\mu}, \ \ \ g_{\mu\nu} \to g_{\mu\nu}, \ \ \ 
\varphi \to \varphi, \ \ \ B_{\mu\nu} \to B_{\mu\nu}, 
\label{tdual}
\end{equation}
where $\Omega$ is an $O(10-D,26-D)$ invariant matrix, {\it i.e.}, 
with the following property:
\begin{equation}
\Omega^T L \Omega = L ,\ \ \ L =\left ( \matrix{0 & I_{10-D}& 0\cr
I_{10-D} & 0& 0 \cr 0 & 0 &  I_{26-D}} \right ),
\label{4dL}
\end{equation}
where $I_n$ denotes the $n\times n$ identity matrix.

At the quantum level, the parameters of $T$-duality
transformation become integer-valued, corresponding to the exact 
symmetry of the perturbative string theory.

\subsection{Explicit Form of the Generating Solution}

In order to obtain the explicit form of the general rotating  
electrically charged solution, we employ the solution generating 
technique, by performing symmetry transformations on a known (neutral)
solution.  Within toroidally compactified heterotic string, an  
approach to obtain the charged solutions from the neutral
one was spelled out in Ref. \cite{SENBH}.  This method was used to 
obtain, {\it e.g.},  general rotating electrically charged 
solutions in four-dimensions \cite{SENBH}, higher dimensional 
general electrically charged static solutions \cite{Peet}  
and rotating solutions  with {\it one} rotational parameter in 
$D$-dimensions \cite{HSEN}, which constitute subsets of 
solutions obtained in this section.

In particular, we perform two $SO(1,1)\subset 
O(11-D,27-D)$ boosts \cite{CYII} on the $D$-dimensional 
Kerr solution, specified by the mass $m$ and $[{{D-1}\over 2}]$ 
angular momenta $l_{1,\cdots, [{{D-1}\over 2}]}$. 
Here $O(11-D,27-D)$ is a symmetry of the effective $(D-1)$-dimensional 
action for stationary solutions of $D$-dimensional heterotic string
compactified on a circle (with the self dual radius 
$R=\sqrt{\alpha^{\prime}}$.  The two $SO(1,1)$ boosts with boost 
parameters $\delta_{1},\ \delta_{2}$ generate the electric charges  
$Q^{(1),(2)}_1$ of the Kaluza-Klein $U(1)$ gauge field  
(associated with the string momentum modes on a circle) 
$A^{(1)\,1}_{\mu}$, and the two-form $U(1)$ gauge field  
(associated with the string winding modes on a circle) 
$A^{(2)}_{\mu\,1}$, respectively.  The solution obtained in
that manner is specified by the ADM mass, {\it two} $U(1)$ charge 
parameters, and $[{{D-1}\over 2}]$ angular momenta 
$J_{1,\cdots, [{{D-1}\over 2}]}$, while the asymptotic values 
of the scalar fields assume canonical values.   Note, however,  
that one can subsequently rescale the asymptotic values 
of the dilaton field and the toroidal moduli.

A subset of $T$-duality transformations, {\it i.e.}, 
$[SO(10-D)\times SO(26-D)]/[SO(9-D)\times SO(25-D)]
\subset O(10-D,26-D)$ transformations, which do not 
affect the $D$-dimensional space-time (and the canonical 
asymptotic values of the scalar fields), provides $(9-D)+(25-D)$ 
additional electric charge parameters, which allow for a general 
rotating electrically charged solution specified by the ADM mass, 
$[{{D-1}\over 2}]$ angular momenta and $36-2D$ electric 
charges, thus consistent with the no-hair theorem
\footnote{A similar analysis is possible in the case of toroidally 
compactified Type IIA string.}.  
Thus, we shall present the generating solution for the most general  
$D$-dimensional rotating electrically charged black holes.

\subsubsection{General Rotating Neutral Solution $-$ Kerr Solution}
The starting point is the  Kerr solutions in $D$-dimensions.  
Following Ref. \cite{MP}, we parameterize it as:
\begin{eqnarray}
ds^2&=&-{{(\Delta-2N)}\over \Delta}dt^2+ {\Delta \over
{\prod^{[{{D-1}\over 2}]}_{i=1}(r^2+l^2_i)-2N}}dr^2
+(r^2+l^2_1\cos^2\theta+K_1\sin^2\theta)d\theta^2 \cr
&+&(r^2+l^2_{i+1}\cos^2\psi_i+K_{i+1}\sin^2\psi_i)
\cos^2\theta\cos^2\psi_1\cdots\cos^2\psi_{i-1}d\psi^2_i \cr
&-&2(l^2_{i+1}-K_{i+1})\cos\theta\sin\theta\cos^2\psi_1\cdots
\cos^2\psi_{i-1}\cos\psi_i\sin\psi_i d\theta d\psi_i \cr
&-&2\sum_{i<j}(l^2_j-K_j)\cos^2\theta\cos^2\psi_1\cdots
\cos^2\psi_{i-1}\cos\psi_i\sin\psi_i\cdots\cos^2\psi_{j-1}
\cos\psi_j\sin\psi_jd\psi_i d\psi_j \cr
&+&{\mu^2_i \over \Delta}[(r^2+l^2_i)\Delta+2l^2_i\mu^2_iN ]
d\phi^2_i - {{4l_i\mu^2_iN}\over \Delta} dtd\phi_i +
\sum_{i<j}{{4l_i l_j\mu^2_i\mu^2_jN}\over \Delta_e}d\phi_i d\phi_j, 
\label{kerr}
\end{eqnarray}
where for 
\begin{itemize}
\item  Even dimensions:
\begin{eqnarray}
\Delta &\equiv& \alpha^2\prod^{{D-2}\over 2}_{i=1}(r^2+l^2_i)
+r^2\sum^{{D-2}\over 2}_{i=1}\mu^2_i(r^2+l^2_1)\cdots(r^2+l^2_{i-1})
(r^2+l^2_{i+1})\cdots(r^2+l^2_{{D-2}\over 2}), \cr
K_i&\equiv&l^2_{i+1}\sin^2\psi_i+\cdots+l^2_{{D-2}\over 2}
\cos^2\psi_i\cdots\cos^2\psi_{{D-6}\over 2}\sin^2\psi_{{D-4}\over 2},
 \ \ \ N=mr ,
\label{edef1}
\end{eqnarray}
and
\begin{eqnarray}
\mu_1 &\equiv&\sin\theta,\ \ \ \  \mu_2\equiv\cos\theta\sin\psi_1,
\ \ \ \ \cdots ,\ \ \ \  
\mu_{{D-2}\over 2}\equiv\cos\theta\cos\psi_1\cdots
\cos\psi_{{D-6}\over 2}\sin\psi_{{D-4}\over 2},\cr
\alpha&\equiv&\cos\theta\cos\psi_1\cdots\cos\psi_{{D-4}\over 2}, 
\label{edef2}
\end{eqnarray}

\item Odd dimensions:

\begin{eqnarray}
\Delta &\equiv& r^2\sum^{{D-1}\over 2}_{i=1}\mu^2_i(r^2+l^2_1)\cdots
(r^2+l^2_{i-1})(r^2+l^2_{i+1})\cdots (r^2+l^2_{{D-1}\over 2}),
\ \ \ N=mr^2, \cr
K_i&\equiv&l^2_{i+1}\sin^2\psi_i+\cdots+l^2_{{D-3}\over 2}
\cos^2\psi_i\cdots\cos^2\psi_{{D-7}\over 2}\sin^2\psi_{{D-5}\over 2}
+l^2_{{D-1}\over 2}\cos^2\psi_i\cdots\cos^2\psi_{{D-5}\over 2}, 
\label{odef1}
\end{eqnarray}
and
\begin{eqnarray}
\mu_1&\equiv&\sin\theta,\ \ \ \  \mu_2\equiv\cos\theta\sin\psi_1,
\ \ \ \  \cdots , \ \ \ \
\mu_{{D-3}\over 2}\equiv\cos\theta\cos\psi_1\cdots
\cos\psi_{{D-7}\over 2}\sin\psi_{{D-5}\over 2},\cr
\mu_{{D-1}\over 2}&\equiv&\cos\theta\cos\psi_1\cdots
\cos\psi_{{D-5}\over 2}.
\label{odef2}
\end{eqnarray}
\end{itemize}
Here, the repeated indices are summed over.  ($i,j$ in
$\psi$  and  $\phi$  run from 1 to $[{{D-3}\over 2}]$  and  
from 1 to $[{{D-1}\over 2}]$, respectively.)

The  Kerr solution is  thus parameterized by the ADM mass $m$ and
$[{{D-1}\over 2}]$ angular momenta $l_{1,\cdots ,[{{D-1}\over 2}]}$.

\subsubsection{Generating Rotating Charged Solution}

The two electric charges $Q_1^{(1)}$ and $Q_1^{(2)}$ necessary 
in parameterizing the generating solution can be induced by imposing 
two $SO(1,1) \subset O(11-D,27-D)$ transformations 
${\bf \Omega}_1$ and ${\bf \Omega}_2$ on the Kerr solution 
(\ref{kerr}).  The boost transformations ${\bf \Omega}_{1,2}$ have 
the following forms:
\begin{eqnarray}
{\bf \Omega}_1&\equiv&\left(\matrix{\cosh\delta_1&\cdot&\cdot&\cdot&
-\sinh\delta_1&\cdot \cr \cdot&I_{9-D}&\cdot&\cdot&\cdot&\cdot \cr
\cdot&\cdot&\cosh\delta_1&\cdot&\cdot&\sinh\delta_1\cr
\cdot&\cdot&\cdot&I_{25-D}&\cdot&\cdot\cr
-\sinh\delta_1&\cdot&\cdot&\cdot&\cosh\delta_1&\cdot\cr
\cdot&\cdot&\sinh\delta_1&\cdot&\cdot&\cosh\delta_1}\right), \cr
{\bf \Omega}_2&\equiv&\left(\matrix{\cosh\delta_2&\cdot&\cdot&
\cdot&\cdot&\sinh\delta_2 \cr
\cdot&I_{9-D}&\cdot&\cdot&\cdot&\cdot \cr
\cdot&\cdot&\cosh\delta_2&\cdot&-\sinh\delta_2&\cdot \cr
\cdot&\cdot&\cdot&I_{25-D}&\cdot&\cdot \cr
\cdot&\cdot&-\sinh\delta_2&\cdot&\cosh\delta_2&\cdot \cr
\sinh\delta_2&\cdot&\cdot&\cdot&\cdot&\cosh\delta_2}\right), 
\label{boosts}
\end{eqnarray}
where the dot denotes the corresponding zero entry.  

The final expressions for the non-trivial $D$-dimensional 
fields take the following forms: 
\begin{eqnarray}
A^{(1)\,1}_t&=&{{N\sinh\delta_1\cosh\delta_1}\over
{2N\sinh^2\delta_1 + \Delta}}, \ \ \ \ \ \ \ \ \ \
A^{(2)}_{t\,1}={{N\sinh\delta_2\cosh\delta_2}\over
{2N\sinh^2\delta_2 + \Delta}},\cr
A^{(1)\,1}_{\phi_i}&=&{{Nl_i\mu^2_i\sinh\delta_1\cosh\delta_2}\over
{2N\sinh^2\delta_1 + \Delta}}, \ \ \ \ \
A^{(2)}_{\phi_i\,1}={{Nl_i\mu^2_i\sinh\delta_2\cosh\delta_1}\over
{2N\sinh^2\delta_2 + \Delta}},\cr
e^{2\varphi}&=&{\Delta^2 \over W},
\ \ \ \ \ \ \ \ \ \ \ \ \ \ \ \ \ \ \ \ \ \ \ \ \ \ \ \
G_{11}={{2N\sinh^2\delta_1 + \Delta}\over
{2N\sinh^2\delta_2 + \Delta}}, \cr
B_{t\phi_i}&=&-{{2Nl_i\mu^2_i\sinh\delta_1\sinh\delta_2
[m(\sinh^2\delta_1+\sinh^2\delta_2)r+\Delta]} \over W}, \cr
B_{\phi_i\phi_j}&=&-4N^2l_il_j\mu^2_i\mu^2_j\sinh\delta_1
\sinh\delta_2\cosh\delta_1\cosh\delta_2[N(\sinh^2\delta_1+
\sinh^2\delta_2) +\Delta]\cr
&\times&[2N^2\sinh^2\delta_1\sinh^2\delta_2
+N\Delta(\sinh^2\delta_1+\sinh^2\delta_2-1)+\Delta^2]/
[(\Delta-2N)W^2],\cr
ds^2&=&\Delta^{{D-4}\over{D-2}} W^{1\over{D-2}}
\left[-{{\Delta-2N}\over W}dt^2+
{dr^2 \over {\prod^{[{{D-1}\over 2}]}_{i=1}(r^2+l^2_i)-2N}}
+{{r^2+l^2_1\cos^2\theta+K_1\sin^2\theta}\over \Delta}
d\theta^2 \right.\cr
&+&{{\cos^2\theta\cos^2\psi_1\cdots\cos^2\psi_{i-1}}
\over \Delta}(r^2+l^2_{i+1}\cos^2\psi_i+K_{i+1}\sin^2\psi_i)
d\psi^2_i \cr
&-&2\sum_{i<j}{{l^2_j-K_j}\over \Delta}
\cos^2\theta\cos^2\psi_1\cdots\cos^2\psi_{i-1}\cos\psi_i\sin\psi_i
\cdots\cos^2\psi_{j-1}\cos\psi_j\sin\psi_j d\psi_i d\psi_j \cr
&-&2{{l^2_{i+1}-K_{i+1}}\over \Delta}
\cos\theta\sin\theta\cos^2\psi_1\cdots\cos^2\psi_{i-1}
\cos\psi_i\sin\psi_i d\theta d\psi_i \cr
&+&{\mu^2_i \over {\Delta W}}[(r^2+l^2_i)\Delta^2
+2Nl^2_i\mu^2_i  +4N^2\sinh^2\delta_1\sinh^2\delta_2
\{r^2+l^2_i(1-\mu^2_i)\} \cr
& &\ \ \ \ +2N\Delta (\sinh^2\delta_1 +
\sinh^2\delta_2)(r^2+l^2_i)]d\phi^2_i - 
{{4Nl_i\mu^2_i\cosh\delta_1\cosh\delta_2}\over W}dtd\phi_i\cr
&+&\left.\sum_{i<j}{{4Nl_il_j\mu^2_i\mu^2_j(\Delta-2N\sinh^2\delta_1
\sinh^2\delta_2)}\over {\Delta W}}d\phi_i d\phi_j\right],
\label{ebh}
\end{eqnarray}
where
\begin{equation}
W \equiv (2N\sinh^2\delta_1 +\Delta)(2N\sinh^2\delta_2 +\Delta)
\label{ebhdef}
\end{equation}
and the other quantities are defined for even and odd dimensions 
in (\ref{edef1}), (\ref{edef2}) and (\ref{odef1}), (\ref{odef2}),  
respectively.

The ADM mass, the angular momenta and electric $U(1)$ charges 
carried by the generating solutions are given by
\footnote{We use the conventions of Ref. \cite{MP}, however, 
we keep in mind that the matter Lagrangian in (\ref{effaction}) 
contains $1/(16\pi G_D)$ prefactor.} 
\begin{eqnarray}
M_{BH}&=&{{\Omega_{D-2}m}\over {8\pi G_D}}
[(D-3)(\cosh^2\delta_1+\cosh^2\delta_2)-(D-4)],\cr
J_i&=&{\Omega_{D-2} \over {4\pi G_D}}ml_i\cosh\delta_1
\cosh\delta_2, \cr
Q^{(1)}_1&=&{{\Omega_{D-2}\over {8\pi G_D}}} (D-3)m\cosh
\delta_1\sinh\delta_1, \cr
Q^{(2)}_1&=&{{\Omega_{D-2}\over {8\pi G_D}}}(D-3)m\cosh
\delta_2\sinh\delta_2,
\label{para}
\end{eqnarray}
where $\Omega_{D-2}\equiv {{2\pi^{{D+1}\over 2}} \over
\Gamma({{D+1}\over 2})}$ is the area of a unit $(D-2)$-sphere 
and $G_D$ is the $D$-dimensional gravitational constant. 
Recall, that for the canonical choice of the asymptotic value 
of the internal metric (toroidal moduli)  $G_{ij}=\delta_{ij}$, 
{\it i.e.}, the compactification is on $(10-D)$ self-dual circles 
with radius $R=\sqrt{\alpha^{\prime}}$, $G_{D}=G_{10}/(2\pi
\sqrt{\alpha^{\prime}})^{10-D}$.  Also, the Kaluza-Klein 
(momentum mode) $Q^{(1)}_1$ and the two-form field (winding mode) 
$Q^{(2)}_1$ charges are quantized in units  
$p/\sqrt{\alpha^{\prime}}$ and $q/\sqrt{\alpha^{\prime}}$, 
respectively.  Here $(p,q)\in {\bf Z}$.

The surface area of the generating solution is given by \cite{HSEN} 
\begin{equation}
A_D=2mr_H\Omega_{D-2}\cosh\delta_1\cosh\delta_2,
\label{area}
\end{equation}
where $r_H$ is the outer horizon determined by 
\begin{equation}
[\prod^{[{{D-1}\over 2}]}_{i=1}(r^2+l^2_i)-2N]_{r=r_H}=0.
\label{horizon}
\end{equation}
Again $N$ is defined in (\ref{edef1}) and (\ref{odef1}) for 
even and odd dimensions $D$, respectively.

The above generating solution has a ring-like singularity 
at $(r,\theta)=(0,{\pi\over 2})$, and therefore the space-time 
structure for the case $m>0$ is that of the Kerr solution. 

With all the angular momenta turned on, the explicit form of 
the thermodynamic entropy, {\it i.e.},  $S_{thermo}\equiv 
A_D/(4G_D)$, (as well as temperature) can be expressed in a  
compact form only for four- and five-dimensional black holes 
\cite{CYIII} and it becomes progressively complicated as the 
dimensionality increases.  In the case of only one rotational 
parameter non-zero, such explicit expressions for the 
entropy and the temperature was obtained in Ref. \cite{HS}.

\subsection{Thermodynamic Entropy of Near-BPS-Saturated Solutions}

The BPS saturated solutions are those whose ADM masses satisfy  
BPS mass formula, and are of special interest.  Such a limit is 
achieved by letting $m\to 0$ and $\delta_i\to\infty$ while 
keeping $me^{2\delta_i}$ 
($i=1,2$) as finite constants.  However, in general with all 
the angular momenta $J_{1,\cdots,[{{D-1}\over 2}]}$ non-zero  
the  BPS-saturated solution has a naked singularity, except in 
the case when at most only one angular momentum is turned on 
(for $D\geq 6$).  The latter property was observed in Ref. 
\cite{HSEN}.  However, even in this special case, the solution 
has a singular horizon, {\it i.e.}, a null singularity.

Therefore, the study of the thermodynamic properties of the 
BPS-saturated rotating black holes is faced with difficulties, 
since the thermodynamic quantities, such as the entropy and 
the temperature are defined at the ``regular'' horizons of 
black holes.   Alternatively, one may want to evaluate such 
quantities at the ``stretched'' horizons as proposed by Sen 
\cite{Sen} for the static electrically charged BPS-saturated 
black holes.  In this case the stretched horizon can be chosen
to be by ${\cal O}(\sqrt{\alpha^{\prime}})$ distance away from 
the singular horizon and the thermodynamic entropy is in 
qualitative agreement (up to a factor of ${\cal O}(1)$) 
\cite{Sen,Peet} with the statistical entropy, {\it i.e.}, the  
logarithm of the degeneracy of BPS-saturated elementary string 
states with the same quantum numbers.   
On the other hand, the area of the rotating BPS-saturated  
black holes, evaluated at the stretched horizon (whose value  
is chosen to be independent of the angular momenta and electric 
charges), turns out to be {\it independent of the angular 
momenta}.  This result is therefore not in accordance with the 
expectations that it should depend on the angular momenta, 
in order to be at least in qualitative agreement with the  
logarithm of the degeneracy of the corresponding rotating 
BPS-saturated string states (see the subsequent Section).  
It may turn out that in order to obtain such an agreement, the 
stretched horizon of rotating BPS-saturated black holes should be 
chosen to depend on the physical parameters
\footnote{We thank A. Sen for a discussion on this point.}.  

Here we would like to propose to consider instead the 
near-BPS-saturated black holes, {\it i.e.}, those with 
non-extremality parameter $m$ small compared with the 
electric charges, and which have regular horizons.  In this 
case, the role of the stretched horizon of the BPS-saturated 
states is traded for the non-extremality parameter $m$ of the 
near-BPS-saturated states. 

The near-BPS-saturated black holes have regular horizons at 
$r_H$ (defined by (\ref{horizon})), provided that the angular 
momentum parameters $l_{1,\cdots,[{{D-1}\over 2}]}$ have the 
magnitude which is smaller than that of $m$.  More precisely,
one has to take $m$ much smaller compared to ${\rm e}^{\delta_i}$, 
such that (when measured in units of $\alpha^{\prime}$) 
$Q_1^{(1),(2)}\gg m={\cal O}(1)$.  In addition, the angular 
momenta are kept small compared to charges, so that (when  
charges are measured in units of $\alpha^{\prime}$) $Q_1^{(1)}
Q_1^{(2)}\gg J^2_{1,\cdots,[{{D-1}\over 2}]}\gg 
{\sqrt{Q_1^{(1)}Q_1^{(2)}}}$.   The first inequality 
ensures the regular horizon, while the second inequality 
ensures that the contribution from the angular momenta to 
the entropy is still non-negligible macroscopically. 

In such a near-BPS-saturated limit, the thermodynamic 
entropy $S_{themro} = {A_D \over {4G_D}}$ can be cast in 
a suggestive  form:
\begin{equation}
S_{thermo}=2\pi \left[{4\over {(D-3)^2}}Q^{(1)}_1
Q^{(2)}_1(2m)^{2\over{D-3}}-{{2}\over{(D-3)}}
\sum_{i=1}^{[{{D-1}\over 2}]} J^2_i\right]^{1\over 2}. 
\label{nexarea}
\end{equation} 

This thermodynamic entropy coincides with the entropy of the 
four- and five-dimensional rotating black holes \cite{CYIII} in 
the near-extreme limit and with two electric charges, only.
 
\section{Near-BPS-Saturated Rotating Electrically Charged Elementary 
String States}

In this section we evaluate the degeneracy of states of 
the (near)-BPS-saturated string states with the {\it same} 
quantum numbers as the near-BPS-saturated rotating black holes.
For the sake of simplicity, we shall address string excitations of 
toroidally compactified heterotic string
\footnote{A related analysis should be possible in the case of 
elementary string excitations of Type IIA string.}.  
The aim is to show that the logarithm of the degeneracy of  
such string states has the same (qualitative) structure as 
the thermodynamic entropy  (\ref{nexarea}) of the 
near-BPS-saturated black hole, thus providing an evidence  
that these string states should serve as microscopic degrees 
of freedom, specifying the microscopic (statistical) origin of  
the black hole thermodynamic  entropy.

The string states specified in terms of the the Kaluza-Klein 
(momentum modes) and two-form field (winding modes) charges
$Q^{(1),(2)}_1$ associated with the NS-NS sector of the  
heterotic string compactification on a self-dual circle with 
radius $R=\sqrt{\alpha^{\prime}}$ have the mass:
\begin{equation}
M_{string}^2= (Q_1^{(1)}+Q_1^{(2)})^2+{4\over {\alpha^{\prime}}}
(N_R-{1\over 2})= (Q_1^{(1)}-Q_1^{(2)})^2+{4\over 
{\alpha^{\prime}}}(N_L-1), 
\label{stringmass}
\end{equation}
where $N_{L,R}$ are the left- and [right-] moving oscillator numbers. 
We would like to determine the degeneracy of states with the mass 
$M_{string}=M_{BH}$ (\ref{para}), and the physical charges 
$Q_1^{(1),(2)}\gg{1\over{\sqrt{\alpha^{\prime}}}}$ (quantized 
in units of $1/\sqrt{\alpha^{\prime}}$)   
and $[{{D-1}\over 2}]$ angular momenta $J_{1,\cdots,
[{{D-1}\over 2}]}\gg 1$ being in the classical regime.  

In the region of $Q_1^{(1),(2)}\gg {1\over{\sqrt{\alpha^{\prime}}}}$,  
the number of the left-moving oscillator modes $N_L$ is large 
and is related to the number of right-moving oscillator modes 
$N_R$ through (\ref{stringmass}) as:
\begin{equation}
N_L= \alpha^{\prime}Q_1Q_2+N_R+{1\over 2},
\label{lmodes}
\end{equation}
For the BPS-saturated states $N_R={1\over 2}$, while for the 
near-BPS-saturated state $N_L\gg N_R > {1\over 2}$ and the 
relationship (\ref{lmodes}) becomes of the form:
\begin{equation}
N_L=\alpha^{\prime} Q_1Q_2+{\cal O}(N_R)\gg N_R,
\label{bpsl}
\end{equation}
and is thus, in the leading order, the same as that for the 
BPS-saturated states.

The leading order contribution to the logarithm of degeneracy 
$d_{N_L}$ of the (near)-BPS-saturated states at the mass 
level $M_{string}$ (\ref{stringmass}) can then be written 
in the form:
\begin{equation}
\log d_{N_L}\sim {2\pi\sqrt{\textstyle{1\over 6} c_{eff}N_L}}=
4\pi\sqrt{N_L}, 
\end{equation}
where  $c_{eff}=26-2=24$. 
 However, this degeneracy of states includes {\it all} the  
((near)-BPS-saturated) states, {\it i.e.}, those with zero and 
non-zero spins, at a particular mass level $M_{string}$.

In order to derive the degeneracy of states at the mass level 
$M_{string}$ and with the particular values of spins 
$J_{1,\cdots,[{{D-1}\over 2}]}$, we employ the conformal field 
theory technique
\footnote{In principle, one should be able to determine the
degeneracy of states by explicitly evaluating the partition 
function with spins included, as initiated in Ref. \cite{RS}.} 
for describing such string states.
Note that the target space quantum numbers of such states 
are described by $[{{D-1}\over 2}]$ angular momenta $J_i$, 
which correspond to the eigenvalues of the $[{{D-1}\over 2}]$ 
$U(1)$ generators of the Cartan sub-algebra of the $O(D-1)$ 
rotational symmetry.  

In the (near)-BPS-saturated limit, these target space quantum 
numbers have a map onto the left-moving world-sheet current 
algebra, specified by $U(1)^{[{{D-1}\over 2}]}$ left-moving 
world-sheet currents $j_i=i\partial_{\bar z}H^i$ 
($i=1,\cdots,[{{D-1}\over 2}]$)
\footnote{Since the right-moving oscillator
modes is $N_R={\cal O}(1)$ in the (near)-BPS-saturated limits, 
the string states have negligible (if any) right-moving  spins.}.   
Consequently, the string states are eigenstates of the 
world-sheet currents whose conformal fields can be written 
in the form:
\begin{equation}
{\bf \Phi}_{J_{1,\cdots,[{{D-1}\over 2}]}} = 
\prod_{i=1}^{[{{D-1}\over 2}]}e^{{iJ_iH^i}}{\bf \Phi}_0,
\label{spinfields}
\end{equation}
with the conformal field ${\bf \Phi}_0$ specifying the 
current-algebra independent part, thus also specifying states 
without target space spins.  The (left-moving) conformal 
dimensions of the two types of conformal fields 
${\bf \Phi}_{J_{1,\cdots,[{{D-1}\over 2}]}}$ 
and ${\bf \Phi}_0$ are related in the following way:
\begin{equation}
{\bar h}_{{\bf \Phi}_{J_{1,\cdots,[{{D-1}\over 2}]}}} = 
{1\over {2}} \sum_{i=1}^{[{{D-1}\over 2}]}
J_i^2 + {\bar h}_{{\bf \Phi}_0}, 
\label{confdim}
\end{equation}
which in turn implies that the degeneracy of states with 
non-zero spins is reduced (in comparison with the degeneracy of 
states with all spins included) precisely by the amount
\footnote{Note that similar argument (within the $D$-brane 
world-sheet current algebra) was used in \cite{BMPV} to 
determine the microscopic entropy of rotating five-dimensional 
BPS saturated black holes.}: 
\begin{equation}
N_L \to {\tilde N}_L= N_L - {1\over {2}} 
\sum_{i=1}^{[{{D-1}\over 2}]}J_i^2,  
\label{modNL}
\end{equation}
which in turn specifies the degeneracy $d_{{\tilde N}_L}$ 
of the (near)-BPS-saturated states at specific mass level 
$M_{string}$ (\ref{stringmass}) and non-zero spins as:
\begin{equation}
S_{stat} \equiv \log d_{{\tilde N}_L} \sim 
4\pi\sqrt{{\tilde N_L}} = 4\pi\left(N_L - {1\over {2}} 
\sum_{i=1}^{[{{D-1}\over 2}]}J_i^2\right)^{1\over 2}. 
\label{statent}
\end{equation}
In order to ensure the statistical nature of the entropy 
we have to maintain $N_L\gg {1\over {2}} 
\sum_{i=1}^{[{{D-1}\over 2}]}J_i^2$, while still allowing 
for the statistically significant contribution from spins, 
{\it i.e.}, $\sqrt{N_L}\sum_{i=1}^{[{{D-1}\over 2}]}
J_i^2\gg 1$.

When $N_L$ is expressed in terms of charges (\ref{bpsl}), 
the statistical entropy assumes the form:
\begin{equation}
S_{stat}=2\pi\left(4\alpha^{\prime}Q_1Q_2-2
\sum_{i=1}^{[{{D-1}\over 2}]}J_i^2\right)^{1\over 2}. 
\label{statentf}
\end{equation}
There is qualitative agreement between the thermodynamic entropy 
(\ref{nexarea}) and the statistical entropy (\ref{statentf}).  
Note however, that since the thermodynamic entropy explicitly 
depends on the non-extremality parameter $m$ , which can be 
determined only up to ${\cal O}(1)$ (in units of 
$\alpha^{\prime}$), the agreement between the two entropies  
can only be qualitative.

\acknowledgments

We would like to thank V. Balasubramanian, C. Hull, F. Larsen,  
A. Tseytlin and especially A. Sen for useful discussions.  
The work is supported by the Institute for Advanced Study 
funds and J. Seward Johnson foundation, U.S. DOE Grant No. 
DOE-EY-76-02-3071, the NATO collaborative research grant CGR 
No. 940870 and the National Science Foundation Career Advancement 
Award No. PHY95-12732.

\end{document}